\documentclass[%
 reprint,
preprintnumbers,
 amsmath,amssymb,
 aps,
prb,
]{revtex4-2}

\usepackage{fancyhdr}
\usepackage{graphicx}% Include figure files
\usepackage{dcolumn}% Align table columns on decimal point
\usepackage[mathlines]{lineno}% Enable numbering of text and display math
\usepackage{isomath}

\usepackage{xcolor}

\newcommand{\usix}{U$_6$Fe }

\begin{document}

\preprint{Draft Version v4}
\pagestyle{plain}
\title{Thermal Equation of State of U$_6$Fe from Experiments and Calculations}
% Force line breaks with \\
\author{Matthew C. Brennan\textsuperscript{1}}
\email[]{mcbrennan@lanl.gov}
\author{Joshua D. Coe\textsuperscript{1}}
\author{Sarah C. Hernandez\textsuperscript{1}} 
\author{Larissa Q. Huston\textsuperscript{1}} 
\author{Sean M. Thomas\textsuperscript{1}}
\author{Scott Crockett\textsuperscript{1}} 
\author{Blake T. Sturtevant\textsuperscript{1}}
\author{Eric D. Bauer\textsuperscript{1}}
\affiliation{\textsuperscript{1}Los Alamos National Laboratory, Los Alamos, New Mexico 87545}
%\affiliation{\textsuperscript{2}Commonwealth Scientific and Industrial Research Organisation, Canberra ACT 2601, Australia}
%\affiliation{\textsuperscript{*}mcbrennan@lanl.gov}

\date{\today}% any date may be explicitly specified

\begin{abstract}
Actinide-bearing intermetallics display unusual electronic, magnetic, and physical properties which arise from the complex behavior of their 5$f$ electron orbitals. Temperature (\textit{T}) effects on actinide intermetallics are well studied, but high pressure (\textit{P}) properties and phase stabilities are known for only a handful of compositions. Furthermore, almost no data exist for simultaneous high \textit{P} and high \textit{T}. We performed ambient-\textit{T} diamond anvil cell X-ray diffraction experiments to study the behavior of the intermetallic \usix upon compression up to 82 GPa. U$_6$Fe remains stable in the tetragonal $I4/mcm$ structure over this pressure range. We also performed ambient \textit{P}, low-\textit{T} diffraction and heat capacity measurements to constrain U$_6$Fe’s thermal behavior. These data were combined with calculations and fitted to a Mie–Grüneisen/Birch–Murnaghan thermal equation of state with the following parameter values at ambient $P$: bulk modulus $B_0$ = 124.0 GPa, pressure derivative $B'_0$ = 5.6, Grüneisen parameter $\Gamma_0$ = 2.028, volume exponent $q$ = 0.934, Debye temperature $\theta_0$ = 175 K, and unit cell volume $V_0 = 554.4$ Å\textsuperscript{3}. We report $T$-dependent thermal expansion coefficients and bond lengths of U$_6$Fe, which demonstrate the anisotropic compressibility and negative thermal expansion of the crystallographic $c$ axis. Additionally, density-functional theory calculations indicate increased delocalization of \usix bonds  at high $P$.
\end{abstract}

\maketitle

%\tableofcontents

\section{\label{sec:level1}Introduction}

The behavior of 5$f$ electrons changes as a function of interatomic spacing over the actinide series, from overlapping and hybridized (3$d$-like) bands in the light actinides to non-bonding and localized (4$f$-like) orbitals in the heavy actinides \cite{Hill1970}. 5$f$ properties become even more diverse when actinides enter intermetallic structures,  which exhibit complex bonding environments with a wide range of interactions both between actinide atoms and between actinides and other elements \cite{Brooks1988,Sechovsky1988}. The bonding and electronic structure of actinide-bearing intermetallics have been extensively investigated, particularly for compounds of U, Np, and Pu, as these elements span the localized/itinerant crossover in 5$f$ behavior and therefore are strongly influenced by their host structure \cite{Kalvius1985,Booth2012,Vitova2017}. 

Because the degree of 5$f$ participation in bonding is so consequential, any property that influences the bonds within an actinide-bearing crystal is likewise significant. Actinide intermetallics include many superconductors (especially of the “heavy fermion” variety), so there has been considerable investigation of their properties at low temperature ($T$), where interatomic distances are smaller and electron overlap and hybridization effects are enhanced \cite{Brodsky1974, Maple1995, Thompson2006}. High pressure ($P$) also decreases interatomic distances and has been shown to significantly alter the electronic and magnetic features of actinide intermetallics \cite{Fournier1985,Potzel1989,Johansson1990,Thompson1993}. Unfortunately, the crystallographic properties underlying these effects are not fully understood, with compressibility and phase stability data for actinide intermetallics being particularly limited. A few equation of state (EoS) studies systematically describe the high $P$ behavior of intermetallic compositions, but these are either restricted to ambient $T$ \cite{dabos1990,Potzel1990,Meresse2000,Shekar2006,Yagoubi2013}, relatively low $P$ \cite{Shukla2020}, or are purely theoretical \cite{Johansson1987,oppeneer2000,Jaroszewicz2013,Schmidt2018,Siddique2019}. This limited body of literature arises from both experimental and computational challenges associated with actinide crystallography. Actinide-bearing materials are intrinsically hazardous and often display low-symmetry structures  that make analysis difficult at high $P$ and $T$. Additionally, computational techniques such as density functional theory (DFT) have been unable to robustly capture the full range of 5$f$ behavior \cite{Pepper1991,Li2001}, though their performance has improved in recent years \cite{Hood2008,Soderlind2018}. In this paper, we combine experiments and calculations to produce a thermal EoS of \usix.

\begin{figure}
\includegraphics[scale = 0.85]{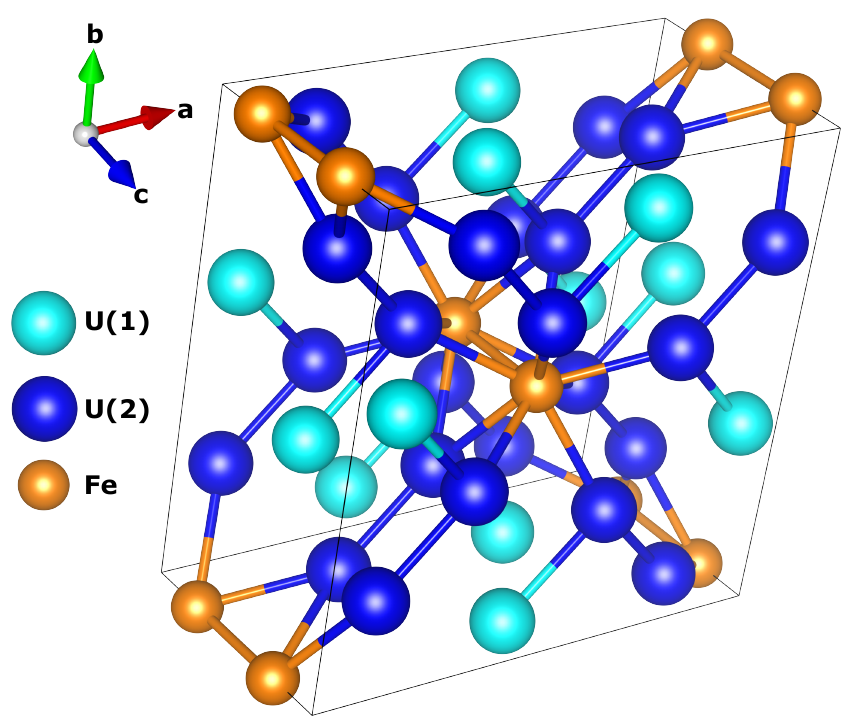}
\caption{\label{Fig. 1} Diagram of the \usix structure showing the shortest bonds between each atomic site within a single unit cell. All U sites lie in $ab$ planes satisfying $z =  n⁄2$, where $z$ is the $c$ coordinate of each such layer and $n$ is an integer. Fe sites lie between these layers and form chains in the $c$ direction. U atoms are more closely packed within layers than between them. Image created with VESTA \cite{Momma2011}.}
\end{figure}

\usix is one of two intermetallics known in the U–Fe system, alongside UFe\textsubscript{2}. It was first synthesized by Manhattan Project chemists \cite{Gordon1949} and recognized as belonging to a group of isostructural peritectic “U\textsubscript{6}X” alloys (where X can be Mn, Co, Fe, Ni, or a pseudobinary combination thereof) by Baenziger et al. \cite{Baenziger1950}. The U\textsubscript{6}X group crystallizes in the “U\textsubscript{6}Mn-type” body-centered tetragonal structure (space group $I4/mcm$, Figure \ref{Fig. 1}), which it shares with Np\textsubscript{6}X and Pu\textsubscript{6}X materials \cite{Brodsky1974}. \usix has been studied in the context of its crystalline-to-amorphous transition \cite{Parkin1986,Huang2016} and its potential as a dispersion fuel in nuclear reactors \cite{Hofman1987,Keiser2003}, but it is most well-known for its superconducting properties. It was the first known superconductor to contain either U or Fe \cite{Chandrasekhar1958}, and remains to this day the U-bearing material with the highest superconducting transition temperature ($T_c \approx 4 $ K). Besides its high $T_c$, \usix displays a variety of unusual superconductivity features, including high-field paramagnetism \cite{DeLong1983}, and high upper critical field values \cite{DeLong1987}, and is often considered an intermediate between heavy fermion materials and other types of superconductors \cite{Yang1989}. 

Like most actinide intermetallics, the electronic and magnetic properties of \usix are interpreted to arise from the bonding environment of its actinide atoms, and thus its internal structure. Within the U\textsubscript{6}X group, $T_c$ is inversely correlated with lattice parameter $a$, with \usix (which has the highest $T_c$) having the smallest $a$ \cite{Engelhardt1975}, and thus the most closely packed U atoms (the closest U--U spacing is 2.66 Å at ambient conditions). Furthermore, $T_c$ slightly increases when U atoms are brought even closer together by increasing $P$~\cite{DeLong1983,DeLong1990,Whitley2016}. These observations imply an important role for structural factors like compressibility and thermal expansion, which control U–U bond lengths but are not well studied. One study reports bond lengths at low $T$~\cite{Kimball1985}, but not the coefficient of thermal expansion ($\alpha$). Another study fits an ambient $T$ EoS to high $P$ \usix data~\cite{Olsen1993}, but improvements in static high $P$ techniques and X-ray detectors over the three decades since then have greatly enhanced the accuracy of EoS determinations. Using updated experimental methods, we extend the \usix EoS to higher $P$ and incorporate measurements of $T$ effects. Combining these with simple physics models allows us to extract the parameters needed for a complete thermal EoS of U$_6$Fe, while DFT simulations help us interpret the effect of $P$ on its electronic structure.

\section{\label{sec:level1}methods}

\subsection{\label{sec:level2}Experimental}
Polycrystalline \usix was prepared by arc-melting a stoichiometric ratio of U and Fe on a water-cooled Cu hearth with a Zr getter. The boule was wrapped in Ta foil and sealed under vacuum, then annealed at 700 °C for 1 month.  A piece of the resulting annealed sample was crushed in a mortar and passed through a 20 \textmu m sieve to create samples appropriate for powder diffraction.  After sieving, the material was  annealed at 400 °C for 24 hours to reduce residual strains from the powdering process. Ambient $P$ powder X-ray diffraction (XRD) patterns were collected using a Malvern Panalytical Empyrean diffractometer in Bragg-Brentano geometry set to 40 mA and 45 kV using Cu K-$\alpha$ radiation. To determine the ambient unit cell volume ($V_0$), a 60-minute exposure was taken at room temperature over the wavevector range 0.4  Å$^{-1} \leq Q \leq$ 5.8  Å$^{-1}$. To determine thermal expansion, a series of 40-minute exposures were taken using an Oxford PheniX cold stage from 12 K to 300 K  over the wavevector range 2.1 Å$^{-1} \leq Q \leq$ 5.8 Å$^{-1}$. Specific heat capacity ($C_P$) measurements were performed in a Quantum Design Physical Properties Measurement System from 2 K to 300 K.

High $P$ powder XRD experiments were performed in a diamond anvil cell (DAC). A 10 \textmu  m diameter flake of the starting material was loaded in a 60 \textmu m diameter sample chamber drilled into a Re gasket pre-indented to 20 \textmu m thickness within a symmetric-type DAC using type Ia anvils with 200 \textmu m diameter culets. In addition to the U$_6$Fe, the sample chamber was loaded with a Cu flake to serve as the pressure standard and Ne gas to serve as the pressure-transmitting medium. Angle-dispersive synchrotron XRD data were collected at Advanced Photon Source Sector 16-BM-D, managed by the High Pressure Collaborative Access Team. The beam energy was 30 keV ($\lambda =$ 0.4133 Å) and the full-width at half maximum beam spot size was 4 × 4 \textmu m$^{2}$.  Diffraction patterns were captured with 30-second exposures on a Pilatus 1M-F detector positioned 211 mm from the sample. A 50 \textmu m diameter pinhole was used to clean up the tails of the beam. The sample $P$ during the experiment was estimated by live pattern integration in Dioptas \cite{Prescher2015} for 0.7  Å$^{-1} \leq Q \leq$ 5.5  Å$^{-1}$ and increased to 82 GPa with an inflating gas membrane \cite{Sinogeikin2015}. 

\subsection{\label{sec:level2}Analysis and Calculations}
Sample $V$ (for ambient $P$ and DAC data) and $P$ (for DAC data) were determined by fitting the positions of 8 \usix and 3 Cu diffraction peaks, respectively, and using these to calculate the lattice parameters of each phase. In the DAC experiment, \usix diffraction peaks overwhelmed those of Cu in patterns focused on the sample flake, making it necessary to collect separate patterns focused on the Cu flake. Using the Cu lattice parameter refined from these patterns, $P$ was calculated using the EoS of Dewaele et al. \cite{Dewaele2004}. Given the weak $P$ gradients present when using Ne as a pressure transmitting medium ($<$ 0.01 GPa/\textmu m at 50 GPa \cite{Klotz2009}), using separate patterns did not significantly affect the determination of $P$ \cite{analysis_supp}.

$C_P$ data at low $T$ were used to determine U$_6$Fe’s $T_c$. The coefficient of electronic specific heat ($\gamma$), and the low-$T$ limit of the ambient-$P$ Debye temperature ($\theta_0$) were determined from a fit to the function
\begin{equation}
    C_P = \gamma T + \frac{12 \pi^4R}{5}  \left( \frac{T}{\theta_0} \right)^3 ,
\end{equation}
where $R$ is the gas constant and the fit was performed over 4–6 K. Low-$T$ XRD data were used to determine the \usix coefficient of thermal expansion ($\alpha$) by the relationship
\begin{equation}
    \alpha_i = \frac{1}{a_i} \left(\frac{\mathrm{d} a_i}{\mathrm{d} T} \right)_P,
\end{equation}
where $a_i$ is unit cell $V$ (or lattice parameter $a$ or $c$) evaluated at constant $P$. High-$P$ XRD data were fit to three isothermal EoS formulations: Murnaghan \cite{Murnaghan1944}, Birch–Murnaghan \cite{Birch1947}, and Rose–Vinet \cite{Vinet1986}. These formulations are, respectively: 
\begin{equation}
    P = \frac{B_0}{B'_0} \left[\left(\frac{V}{V_0}\right)^{-B'_0} -1\right],
\label{M}
\end{equation}
\begin{equation}
    P = \frac{3 B_0}{2} \left(\eta^{-7} - \eta^{-5} \right)   \left[1 + \frac{3}{4} \left(B'_0-4\right) \left(\eta^{-2} - 1\right)\right],
\label{BM}
\end{equation} and
\begin{equation}
    P = 3 B_0 \left(\frac{1-\eta}{\eta^{2}}\right)\exp\left[{\frac{3}{2}(B'_0-1)(1-\eta)}\right]
\label{RV}
\end{equation}
where $B_0$ is the bulk modulus at ambient $P$, $B'_0$ is its pressure derivative, and $\eta = (V/V_0)^{1/3}$.

A tabular EoS was constructed based on the SESAME framework \cite{Lyon1992}, which decomposes the Helmholtz free energy ($F$) into its cold (0 K), thermal ionic, and thermal electronic contributions. Expressed in terms of specific density ($\rho$),
\begin{equation}
    F(\rho,T)=\phi(\rho)+F_{\rm{ion}}(\rho,T)+F_{\rm{elec}}(\rho,T) .
\end{equation}
Identical decompositions apply to the internal energy ($E$) and $P$. The cold contribution ($\phi$) was based on a Birch--Murnaghan fit to the DAC data reported below. The ionic model parameters were adjusted in combination with the cold values of $\rho$, $B$, and $B'$ to recover the experimental values $\rho_0$, $B_0$, and $B'_0$ reported below, where the subscript ‘0’ indicates the ambient reference state. $F_{\rm{ion}}$  was based on a standard Debye model \cite{McQuarrie2000} in the quasiharmonic approximation \cite{Anderson1995}, where $\theta_0$ was adjusted to match $C_P$ data from 0–300 K. The density-dependence of $\theta$ was governed by that of the Grüneisen parameter ($\Gamma$),

\begin{equation}
\Gamma(\rho) =
\begin{cases}
   \Gamma(\infty) + c_1 \left(\frac{\rho_0}{\rho}\right) + c_2 \left(\frac{\rho_0}{\rho}\right)^{2} & \text{, } \rho \geq \rho_0 \\
   \Gamma(0) + c_3 \left(\frac{\rho}{\rho_0}\right) + c_4 \left(\frac{\rho}{\rho_0}\right)^{2} & \text{, } \rho \leq \rho_0
  \end{cases}
  \end{equation}

where
\begin{equation}
    \Gamma = \frac{d \ln \theta}{d \ln \rho} .
\end{equation}
The coefficients $c_1-c_4$ were adjusted automatically to maintain continuity of $\Gamma$ and its first derivative ($\Gamma'$) at $\rho_0$, while the values $\Gamma(0)=2/3$ and $\Gamma(\infty)=1/2$ were set to those of the monatomic ideal gas and the one-component plasma \cite{Nagara1985,Petrov1994}, respectively. $\Gamma_0= 2.06$ to recover the thermal expansion results reported below and $\Gamma'_0 = -\Gamma_0$ by convention. The electronic contribution was based on Thomas-Fermi-Dirac theory \cite{Feynman1949,Cowan1957}, whose only input is the mean atomic number $\bar{Z} = 82.571$.

\usix cold energy and electronic structure were calculated based on DFT with the Perdew-Burke-Ernzerhof exchange-correlation functional \cite{Perdew1996}. Calculations were performed using VASP \cite{Kresse1993,Kresse1994,Kresse1996a,Kresse1996b}, with no spin-orbit-coupling or spin-polarization applied. VASP cold curve calculations used a 500 eV kinetic energy plane-wave cutoff, a 2 x 2 x 4 k-point grid, an energy convergence of $1\times 10^{-5}$eV, and ions were relaxed until the Hellmann-Feynman forces \cite{Feynman1939} on each were $<$0.01 eV/Å. A sequence of volumes was generated by uniform contraction or expansion of $a$, $c$, and the internal degrees of freedom about a reference structure. At each $V$, all lattice parameters and degrees of freedom were relaxed to minimize the internal energy subject to the constraint that volume be preserved. This yielded an $E(V)$ locus that was fit to the volume integral of Eq.~(\ref{BM}),
\begin{equation}
    E = E_0 + \frac{16 V_0 B_0}{9} [ (\eta^{-2}-1)^{3}B'_0+(\eta^{-2}-1)^{2}(6-4\eta^{-2})]
\end{equation}
and then comparison to experiment was made by evaluating Eq.~(\ref{BM}) based on the $V_0$, $B_0$, and $B'_0$ that resulted. 

\section{\label{sec:level1}Results}

\begin{figure}
\includegraphics{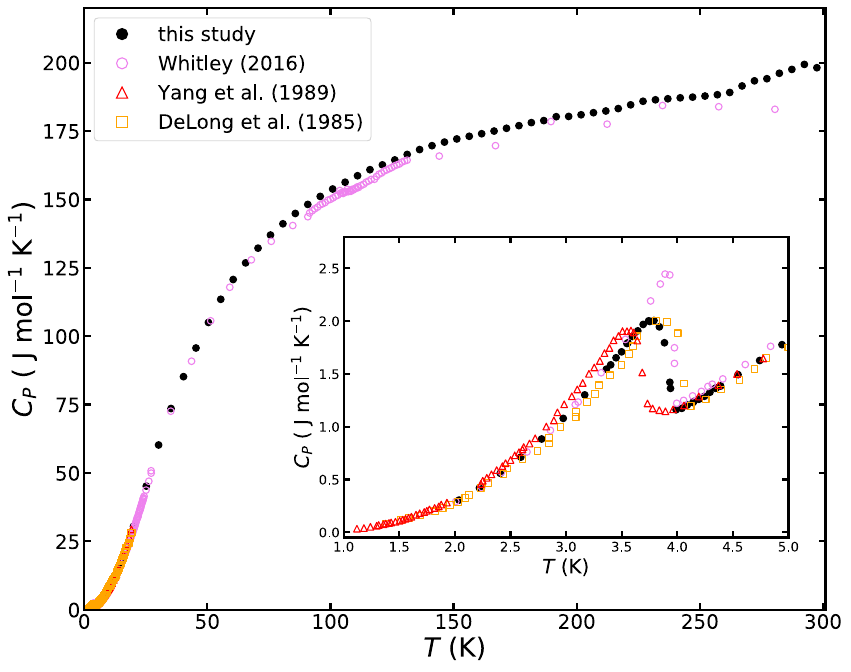}
\caption{\label{Fig. 2} Measured specific heat ($C_P$), along with  literature data~\cite{DeLong1985,Yang1989,Whitley2016}. Inset: $C_P$ vs $T$ below 5 K. The discontinuity near $T$=4 K indicates the superconducting transition.}
\end{figure}

Data from our low-$T$ heat capacity measurements are shown in Figure \ref{Fig. 2}. These results generally reproduce previous measurements and indicate a superconducting transition $T_c \approx 4$ K \cite{Chandrasekhar1958} with $\Delta C_P/\gamma T_c = 1.3$. We did not observe any indications of the structural or charge density wave state transition that has been suggested to occur near 100 K \cite{DeLong1985,Kimball1985,Lemon1987,Whitley2016}, although our data may be insufficiently dense to detect subtle features in this region. 

\begin{figure}
\includegraphics[scale=0.8]{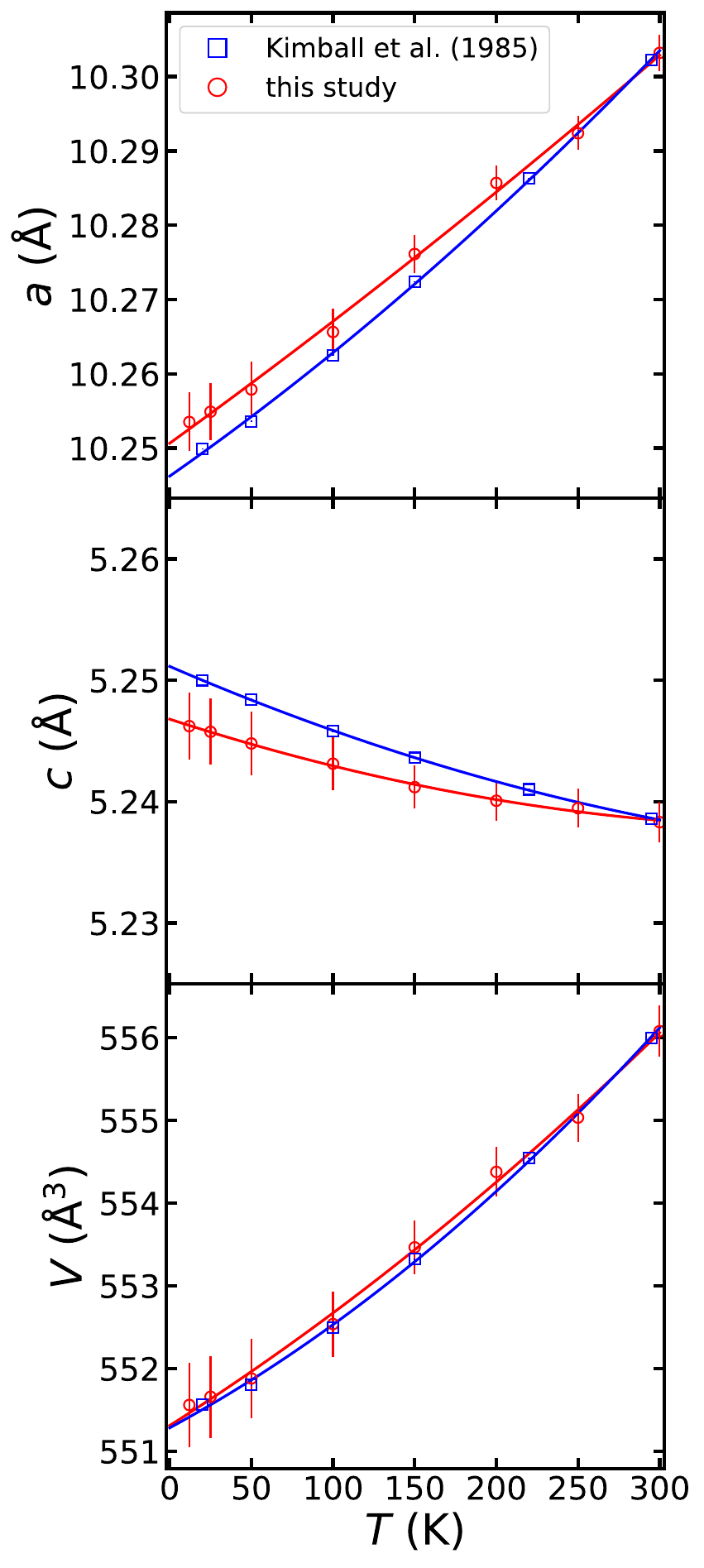}
\caption{\label{Fig. 3} Variation of unit cell parameters $a$ (top), $c$ (center), and volume (bottom) with $T$. Trendlines shown are quadratic fits to the data.}
\end{figure}

Ambient-$P$, low-$T$ XRD measurements are shown in Figure \ref{Fig. 3} and reproduce the unusual anisotropic thermal expansion previously reported in \usix \cite{Kimball1985,Vaishnava1985}. As $T$ decreases, \usix contracts along the $a$ axis but expands along the $c$ axis; this orientation-dependent thermal effect is also observed at low $T$ in other U-bearing and heavy fermion materials and is attributed to enhanced electronic effects in these systems \cite{Barrera2005}. Table \ref{tab:table1} lists the volumetric and directional coefficients of thermal expansion as a function of $T$, calculated from the fits shown in Figure \ref{Fig. 3} \cite{alpha_supp}. 

\begin{table}[b]
\caption{\label{tab:table1} Parameters for thermal expansion coefficients (units of 1/K) as a function of $T$. Parameter $x_1$ has units of 1/K, $x_2$ has units of 1/K$^2$, and $x_3$ has units of 1/K$^3$, such that $\alpha = x_1 + x_2T +x_3T^2$. Note that Kimball et al. \cite{Kimball1985} does not explicitly report $\alpha$; values from that study were calculated from reported structural parameters.
}
\begin{ruledtabular}
\begin{tabular}{llccc} %1 left, 3 centered
\textrm{}&
\textrm{}&
\textrm{$x_1 (\times 10^{-6})$}&
\textrm{$x_2 (\times 10^{-9}$)}&
\textrm{$x_3 (\times 10^{-12}$)}\\
\colrule
$\alpha_a$&(this study) & 15.522(2) & 9.51(3) & -0.25(9)\\
&(Kimball et al.) & 15.8(6) & 10.3(1) & -11(3)\\
\hline
$\alpha_c$&(this study) & -8.365(5) & 20.37(5) & 0.14(1)\\
&(Kimball et al.) & -7.6(2) & 26(2) & -27(5)\\
\hline
$\alpha_V$&(this study) & 22.663(5) & 39.76(8) & -1.7(3)\\
&(Kimball et al.) & 23.8(3) & 55(4) & -67(9)
\end{tabular}
\end{ruledtabular}
\end{table}

\begin{figure*}
\includegraphics[scale = 1.0] {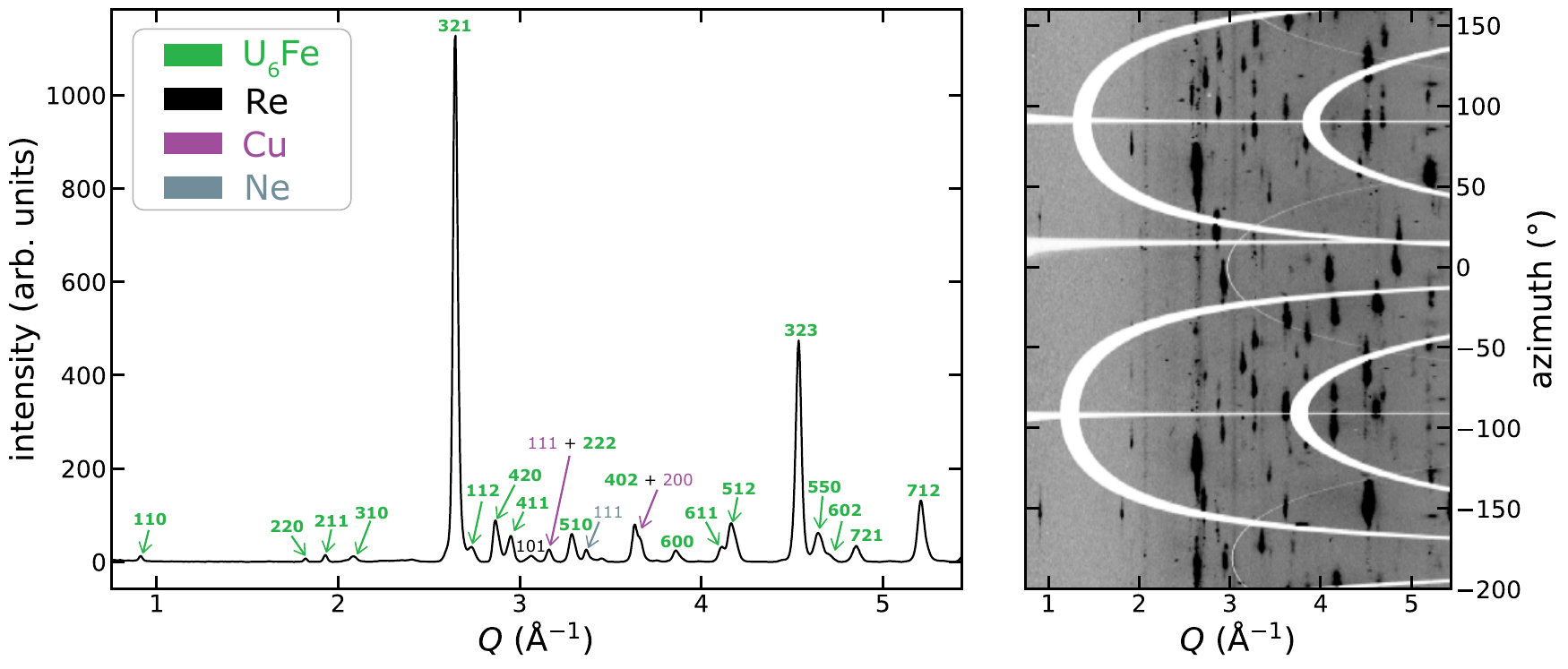}
\caption{\label{Fig. 4} An integrated X-ray diffraction pattern (left) and its corresponding caked 2D diffraction image (right). Reflections are labeled with their corresponding Miller indices and color coded by phase (\usix reflections are also labelled with bold text). Intensity variations in the diffraction rings indicate a degree of preferred orientation in our sample.}
\end{figure*}

A representative XRD pattern from our DAC experiment is shown in Figure \ref{Fig. 4}. This pattern was collected at 36 GPa and shows peaks from the \usix sample, the Cu pressure standard, the Ne pressure transmitting medium, and the Re gasket, although the sample  is by far the strongest diffractor. Despite the high signal-to-background ratio we did not observe \usix $002$ (which should be near the left shoulder of \usix $321$) in the high-$P$ patterns. This absence is likely due to the preferred orientation (i.e., texture) visible in our diffraction patterns. The \usix unit cell parameters were determined as a function of $P$ up to 82 GPa \cite{unitcell_supp}. \usix maintains the body-centered tetragonal ($I4/mcm$) structure over this entire pressure range, consistent with the absence of a room-temperature phase change in pure U until much higher $P$ \cite{Kruglov2019}. 

\section{\label{sec:level1}discussion}
\begin{figure}
\includegraphics[scale=1.0]{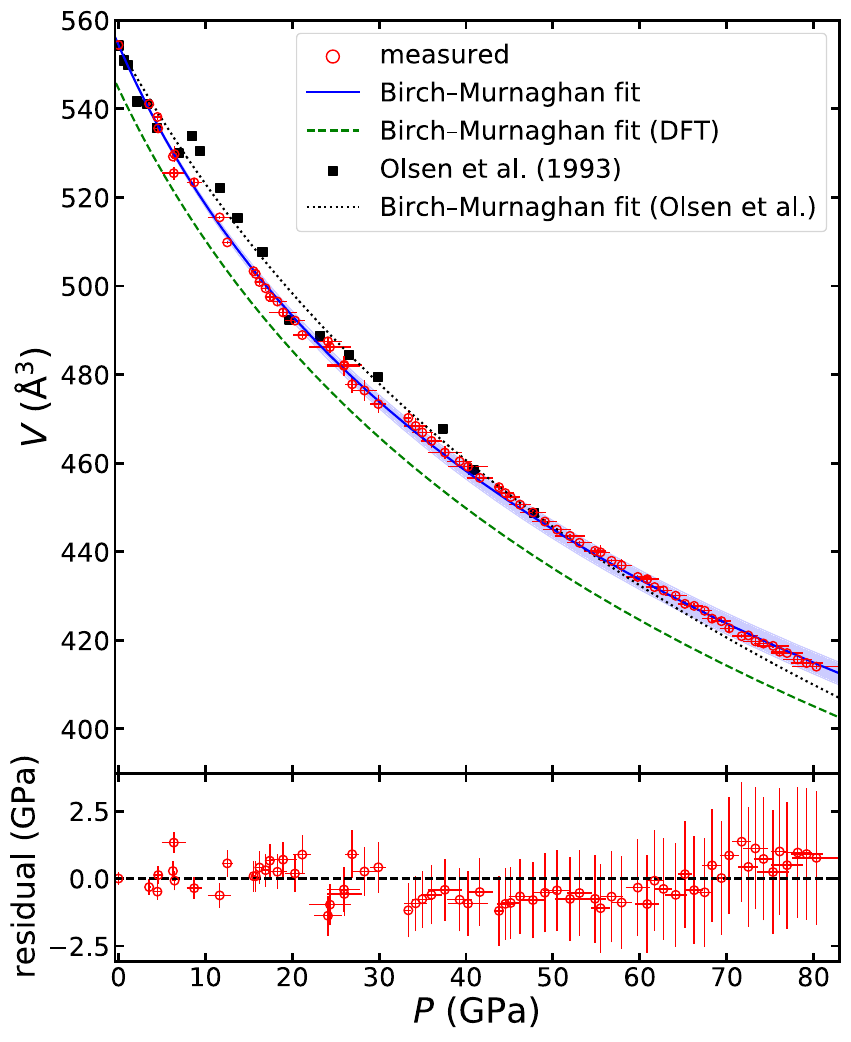}
\caption{\label{Fig. 5} Unit cell $V$ measured in this study, reported by Olsen et al. \cite{Olsen1993}, unit cell $V$ calculated with DFT, and corresponding Birch–Murnaghan EoS fits. The shaded blue region is a $2\sigma$ error envelope calculated from the uncertainty of the fit parameters (Table II). The lower panel shows residuals quantified as the difference between Birch–Murnaghan calculated $P$ and observed $P$ for each of the measured $V$. }
\end{figure}

Figure \ref{Fig. 5} shows the change in unit cell $V$ as $P$ increases at ambient $T$, and Table \ref{tab:EoStable} lists EoS parameters fit with $V_0$ fixed at our measured ambient-$P$ value. Treating $V_0$ as a free parameter in the fit does not change any of the parameter values within uncertainty, although it does increase the $B_0$, $B'_0$ covariance by a factor of 4. Compared to values fit to the data of Olsen et al. \cite{Olsen1993,olsen_supp}, we find that \usix is more compressible (i.e., $B_0$ is lower), with the compressibility increasing more rapidly as $P$ increases (i.e., $B'_0$ is higher). Compared with U$_6$Fe, $\alpha$-U is slightly more compressible ($B_0$ = 114.5 GPa, $B’_0$ = 5.46) \cite{Dewaele2013}, while UFe\textsubscript{2} is much less compressible ($B_0$ = 239 GPa, $B’_0$ = 3) \cite{Itie1986}. This is unexpected since UX\textsubscript{2} intermetallics are thought to have localized $f$ electrons that do not participate in bonding and should therefore have lower bulk moduli \cite{Potzel1990,Shekar2006}. Future measurements of materials within the U\textsubscript{6}X, Np\textsubscript{6}X, or Pu\textsubscript{6}X groups would allow for a comparison of compressibility as a function of actinide packing within the U\textsubscript{6}Mn-type structure. 

\begin{table*}
\caption{\label{tab:EoStable} Isothermal EoS parameters for U$_6$Fe. Parameters marked ``this study" are derived from DAC measurements with $V_0$ fixed at the measured 0 GPa value and $B_0$ and $B'_0$ fit by orthogonal distance regression to the data shown in Figure \ref{Fig. 5}.
}
\begin{ruledtabular}
\begin{tabular}{lccccc} %1 left, 3 centered
Fitting Form & Data Source & $V_0$ (Å$^3$) & $B_0$ (GPa) & $B'_0$ & cov($B_0$,$B'_0$) \\
\colrule
Rose–Vinet & this study & 554.4(10) & 122.9(11) & 5.8(1) & -0.0535 \\
\hline
Murnaghan & this study & 554.4(10) & 127.3(11) & 4.9(3) & -0.0613\\
\hline
Birch–Murnaghan & this study & 554.4(10) & 124.0(12) & 5.6(1) & -0.0364 \\
& DFT (0 K) & 539.0(1) & 137.9(4) & 4.8(0) & -0.0106 \\
& DFT (300 K) & 544.6 & 131.3 & 4.86 & -- \\
& Olsen et al. & 555.8(29) & 155.4(85) & 3.6(6) & -3.05
\end{tabular}
\end{ruledtabular}
\end{table*}

Table~\ref{tab:EoStable} also lists parameters based on fits to the DFT cold curve. Because the DFT calculations are at 0 K, comparison of those results with DAC data requires that the former be ``corrected" to room temperature based on $F_{\rm{ion}}$. Uncertainties in the 0 K parameters are small, while those in the 300 K values are more difficult to estimate. Consistency of theory and experiment is good: DFT predicts $\sim 2$\% lower $V_0$ and $\sim 6$\% higher $B_0$.

We used the results of our SESAME calculations to produce a Mie-Grüneisen thermal EoS of the form
\begin{equation}
    P(V,T) = P_{iso} +  P_{thermal}
\end{equation}
where $P_{iso}$ is the isothermal pressure at 300 K (Equations 3–5) and $P_{thermal}$ is the thermal pressure using a quasiharmonic approximation \cite{Anderson1995}:
\begin{equation}
    P_{thermal} = \frac{9nR\Gamma}{V}\left[\frac{\theta}{8}+T\left(\frac{T}{\theta}\right)^{3} \int_{0}^{\theta / T} \frac{z^3 dz}{e^z -1}\right]
\end{equation}
where $\Gamma$ and $\theta$ are the Grüneisen parameter and Debye temperature, respectively. These are evaluated with reference to their ambient values ($\Gamma_0$ and $\theta_0$) by:
\begin{equation}
    \Gamma = \Gamma_0 \left(\frac{V}{V_0}\right)^q
\end{equation} and
\begin{equation}
    \theta = \theta_0 e^{(\Gamma - \Gamma_0)/q}
\end{equation}
where $q$ is a dimensionless fitting parameter. Parameter values are listed in Table \ref{tab:table3}. Note that we report both the limit of $\theta_0$ as $T \rightarrow 0$ K and a more broadly applicable  $\theta_0$ fit over the entire range of our $C_P$ data. 

\begin{table*}
\caption{\label{tab:table3}Thermal parameters ($\gamma$: electronic specific heat coefficient, $\theta_0$: Debye temperature, $T_c$: superconducting transition temperature, $\Gamma_0$: Grüneisen parameter, $q$: Grüneisen volume exponent) compared to values from previous studies \cite{DeLong1985,Kimball1985,Yang1989,Whitley2016}. $\gamma$ and the low-$T$ limit of $\theta_0$ were determined by fitting specific heat data from 4–6 K. }
\begin{ruledtabular}
\begin{tabular}{lccccc}
 &this study& Whitley (2016) & Yang et al. (1989) & DeLong et al. (1985) & Kimball et al. (1985)\\
\\ \hline
 $\gamma$ (mJ mol$^{-1}$ K$^{-2}$) & 158(20) & 154(25) & 151.7 & 150(3) &\\
$\theta_0 (T\rightarrow0)$(K) & 118.0(1) & 113 & 114.3 & 116 &\\
$\theta_0$ (K) & 175 & & & & 125\\
$T_c$ (K) & 3.9(1) & 4 & 3.695(1) & 3.9 & 3.76(17)\\
$\Gamma_0$ & 2.208 & & & & \\
$q$ & 0.934(1) & & & & 
\end{tabular}
\end{ruledtabular}
\end{table*}

\begin{figure}
\includegraphics[scale=0.85]{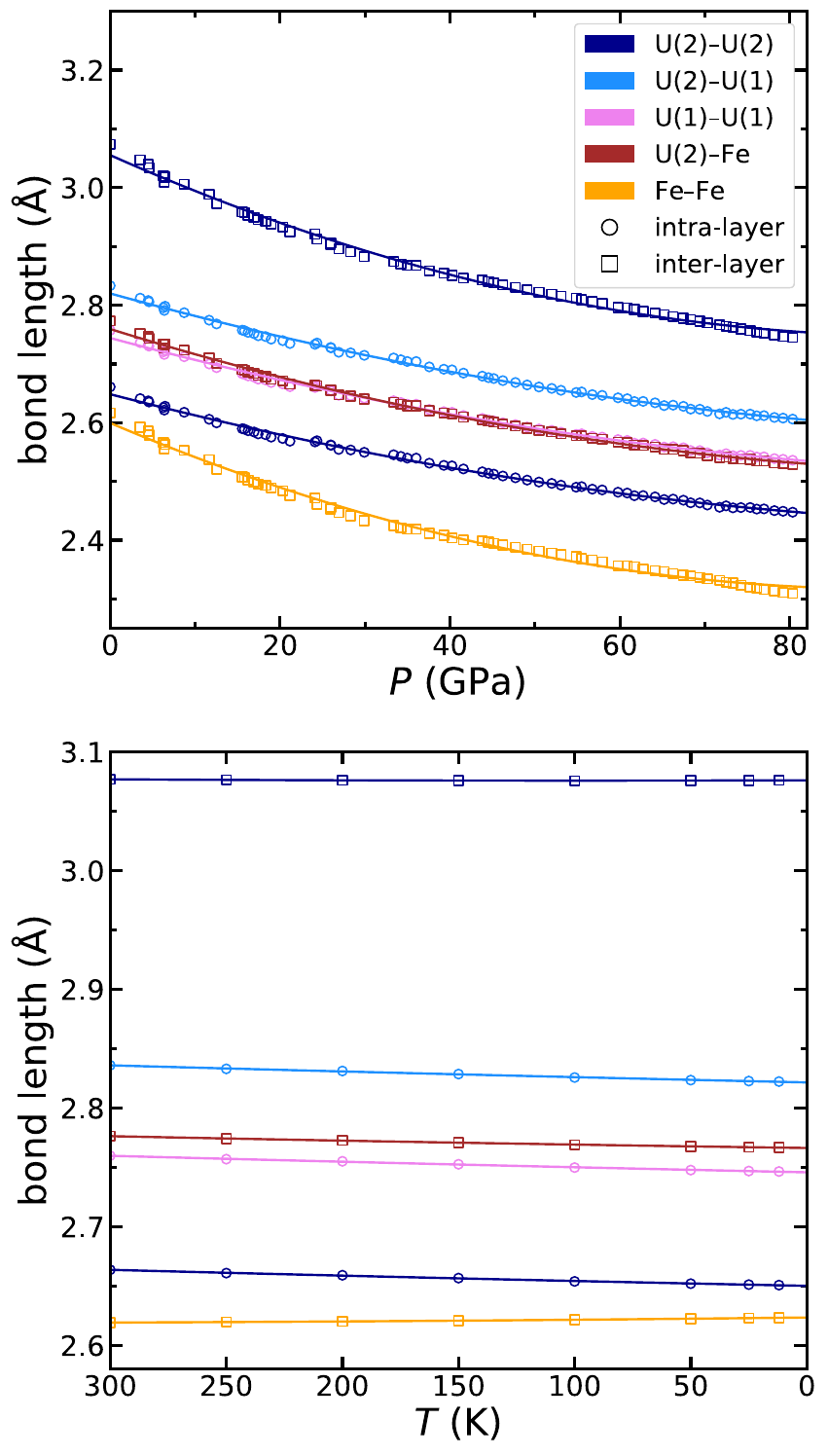}
\caption{\label{Fig. 6} Variation of bond lengths in the \usix structure with $P$ at ambient $T$ (top panel) and $T$ at ambient $P$ (bottom panel). The nearest-neighbor distance between each atomic site is shown, plus the next-nearest-neighbor distance for U(2)–U(2). “Intra-layer” bonds lie in the $ab$ plane, while “inter-layer” bonds have a $c$ component (see Figure \ref{Fig. 1}).} 
%U(1)–U(1) bonds are  almost identical in length to U(2)–Fe and are omitted for clarity.}
\end{figure}

Bond lengths for our XRD experiments were calculated with Mercury \cite{Macrae2020} and are shown in Figure \ref{Fig. 6}. U atoms are most densely packed within their layers (Figure \ref{Fig. 1}), meaning that the shortest bonds between them are all “intra-layer” and lie entirely in the $ab$ plane. Since the $c$ axis is the most compressible crystallographic direction \cite{unitcell_supp}, these intralayer bonds compress more slowly than the next-nearest-neighbor “inter-layer” U--U bonds, which have a $c$ component. At very high $P$, inter-layer bonds may become comparable to the length of intra-layer bonds, possibly changing the coordination of U and leading to a change in crystal structure. The most compressible bonds are those of the Fe–Fe chains, which lie entirely in the $c$ direction. Contrary to their high $P$ behavior, Fe–Fe bonds lengthen at low $T$ due to the anisotropic negative thermal expansion mentioned above. $\alpha$-U similarly displays higher compressibility in the direction of negative thermal expansion \cite{Dewaele2013}, but this behavior is unusual in U-bearing intermetallics. Intermetallics are generally both more thermally expansive and more compressible in the U–U bond direction \cite{Maskova2012}. 

\begin{figure}
    \includegraphics[scale=1.0]{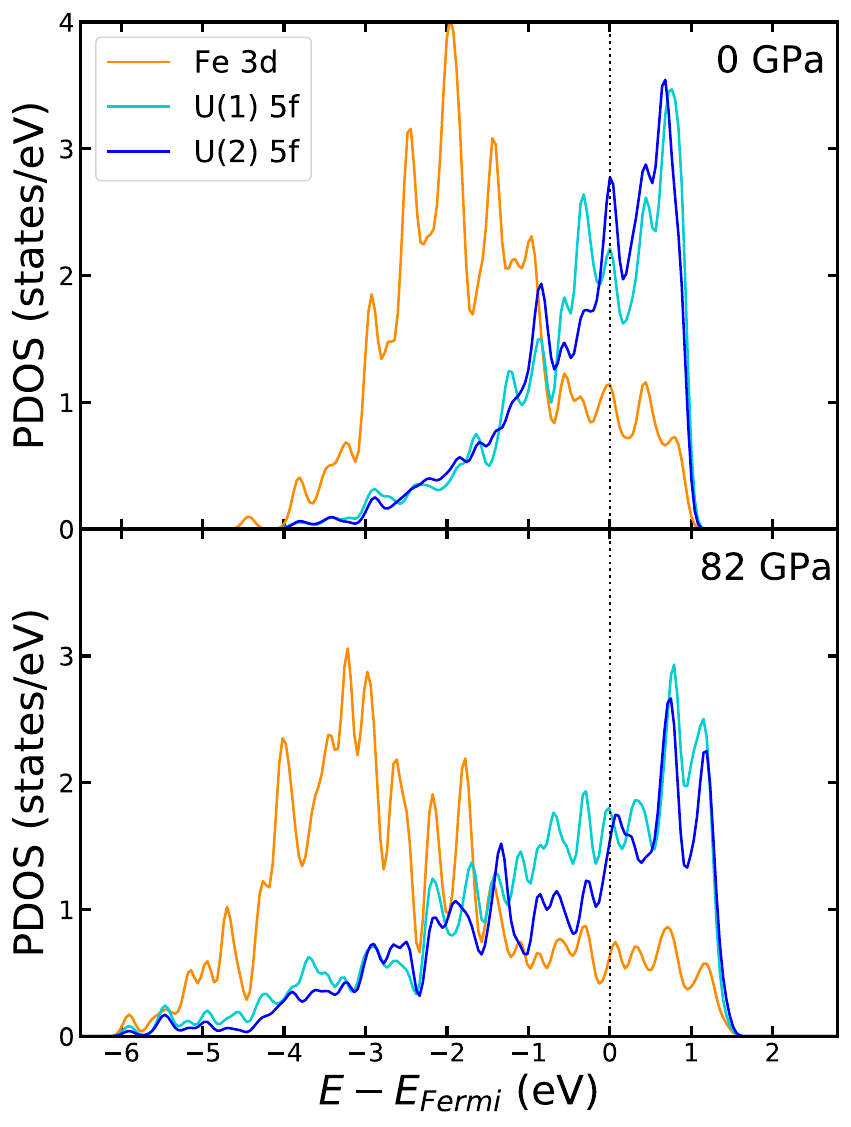}
    \caption{Partial density of states of selected orbitals as a function of energy. There is clear evidence of spectral broadening (and thus increased delocalization) over our compression range.}
    \label{fig:Fig. 7}
\end{figure}

The effect of cooling from 300 K to $T_c$ on U–U bond length is approximately the same as the effect of compressing to 1 GPa. This may be related to the observed increase in $T_c$ upon compression at low $P$ \cite{DeLong1990,Whitley2016} since the superconductivity of U\textsubscript{6}X materials depends on the degree of U orbital hybridization \cite{Hill1968}. On the other hand, the partial density of states (PDOS) for DFT-based unit cells did not demonstrate significant changes in the distribution of electronic states at low $P$. Over a much larger range of compression (0--82 GPa), a continuous increase in the delocalization of both U 5$f$ and Fe 3$d$ electrons is implied by the broadening of their respective PDOS (Figure \ref{fig:Fig. 7}). As previously predicted for light actinides, compressing \usix increases the U atoms' 5$f$ occupancy \cite{Soderlind2011} from 2.9 electrons at ambient $P$ to 3.6 at 82 GPa.

\section{\label{sec:level1}conclusions}
We have conducted low $T$ measurements of heat capacity and thermal expansion and high $P$ measurements of unit cell volume in U$_6$Fe. These data were used to construct a tabular SESAME EoS and a closed-form Mie–Grüneisen/Birch–Murnaghan EoS with parameters $B_0$ = 124.0 GPa, $B’_0$ = 5.6, $\Gamma_0$ = 2.028, $\theta_0$ = 175 K, $q$ = 0.934 and $V_0$ = 554.4 Å$^3$. This approach may be of use for other materials for which it is impractical to collect simultaneous high $P$ and $T$ data. DFT calculations are consistent with experimental parameterizations and indicate that \usix bonds are increasingly delocalized at high $P$. These results contribute to our understanding of \usix crystallography; similar investigations of other U\textsubscript{6}Mn-structured compounds will help further explore the role of 5$f$ bonds in determining the macroscopic properties of actinide-bearing intermetallics. 

%\clearpage
\begin{acknowledgments}
  M. C. Brennan was supported in part by a fellowship from the Glenn T. Seaborg Institute. Portions of this work were performed at HPCAT (Sector 16), Advanced Photon Source, Argonne National Laboratory. HPCAT operations are supported by DOE-NNSA’s Office of Experimental Sciences. The Advanced Photon Source is a U.S. Department of Energy Office of Science User Facility by Argonne National Laboratory under Contract No. DE-AC02-06CH11357.This work was supported by the U.S. Department of Energy through the Los Alamos National Laboratory. Los Alamos National Laboratory is operated by Triad National Security, LLC, for the National Nuclear Security Administration of the U.S. Department of Energy (Contract No. 89233218CNA000001).  This work was supported in part by the U.S. Department of Energy's Dynamic Material Properties Program and Advanced Simulation and Computing Program.
\end{acknowledgments}

%\nocite{*}

\clearpage
\bibliography{U6Fe.bib}% Produces the bibliography via BibTeX.

\end{document}

\preprint{Draft Version v4}

\title{Supplemental Material for ``Thermal Equation of State of \usix from Experiments and Calculations''}% Force line breaks with \\

\author{Matthew C. Brennan\textsuperscript{1}}
\email[]{mcbrennan@lanl.gov}
\author{Joshua D. Coe\textsuperscript{1}}
\author{Scott C. Hernandez\textsuperscript{1}} 
\author{Larissa Q. Huston\textsuperscript{1}} 
\author{Sean M. Thomas\textsuperscript{1}}
\author{Scott Crockett\textsuperscript{1}} 
\author{Blake T. Sturtevant\textsuperscript{1}}
\author{Eric D. Bauer\textsuperscript{1}}
\affiliation{\textsuperscript{1}Los Alamos National Laboratory, Los Alamos, New Mexico 87545}
%\affiliation{\textsuperscript{2}Commonwealth Scientific and Industrial Research Organisation, Canberra ACT 2601, Australia}
%\affiliation{\textsuperscript{*}mcbrennan@lanl.gov}

\date{\today}% any date may be explicitly specified

\maketitle
\tableofcontents
\clearpage

\section{\label{sec:level1}Thermal expansivity}

\begin{figure}[h]
\centering
\includegraphics[scale = 0.5]{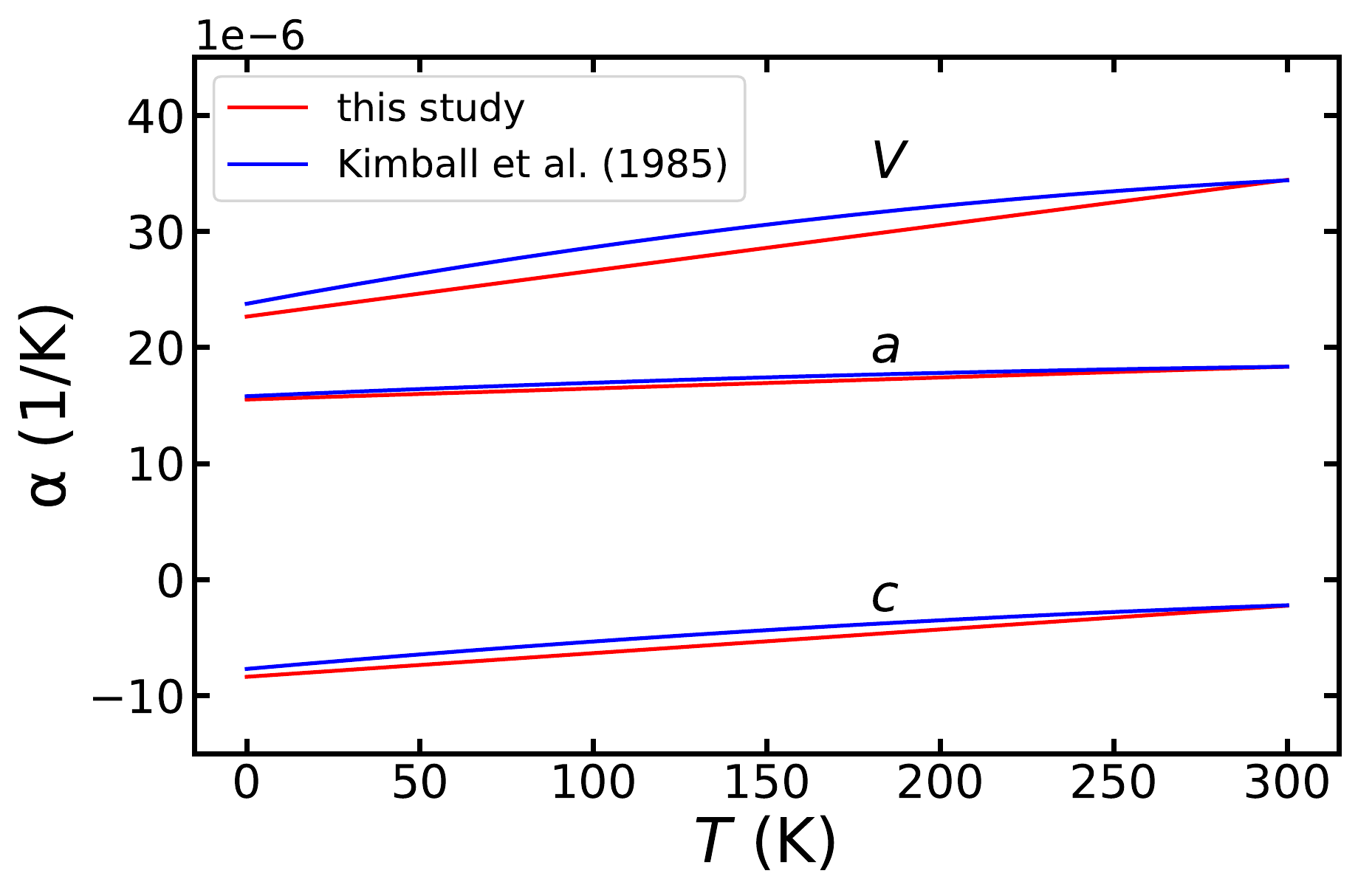}
\caption{\label{Fig. 1} Variation of the thermal expansion coefficient for the unit cell dimensions ($a$, $c$) and the unit cell volume ($V$). See also Figure 3 and Table 1 in the main text.}
\end{figure}

\section{\label{sec:level1}Determining $V$ and $P$ in DAC experiments}
\begin{figure}[H]
   \centering
    \includegraphics[scale = 0.3]{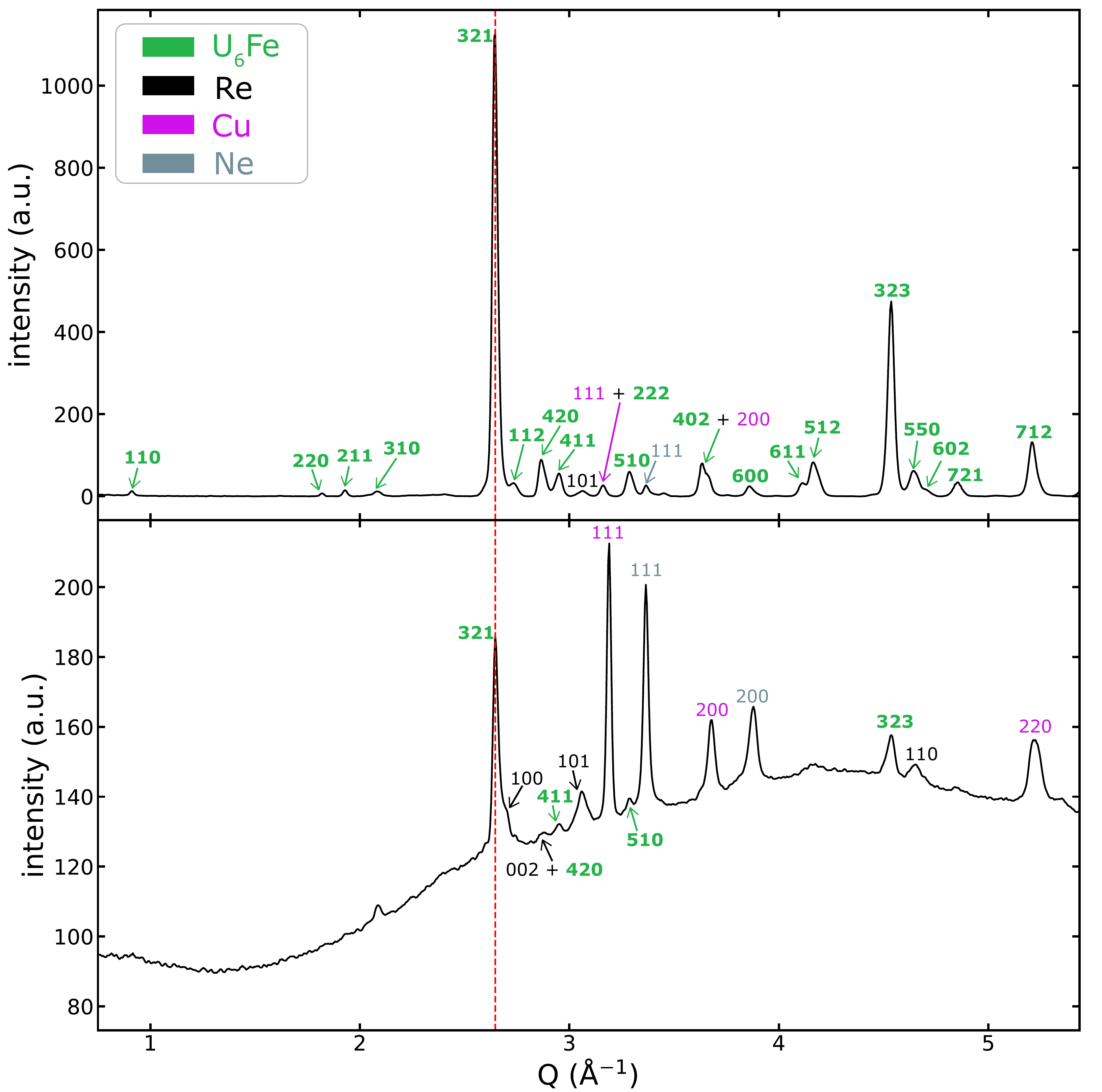}
    \caption{An integrated diffraction pattern from focused on the \usix sample (top) and Cu pressure standard (bottom) at the same membrane pressure. Reflections are labelled with their corresponding Miller indices and color coded by phase (\usix reflections are also labelled with bold text). At a given membrane pressure, sample $V$ was determined from \usix peaks and sample $P$ was determined from Cu peaks in their respective patterns. The dashed line shows the consistency in the \usix $321$ position between the two patterns.}
    \label{fig:patternfig}
\end{figure}

\begin{figure}[H]
    \centering
    \includegraphics[scale = 0.5]{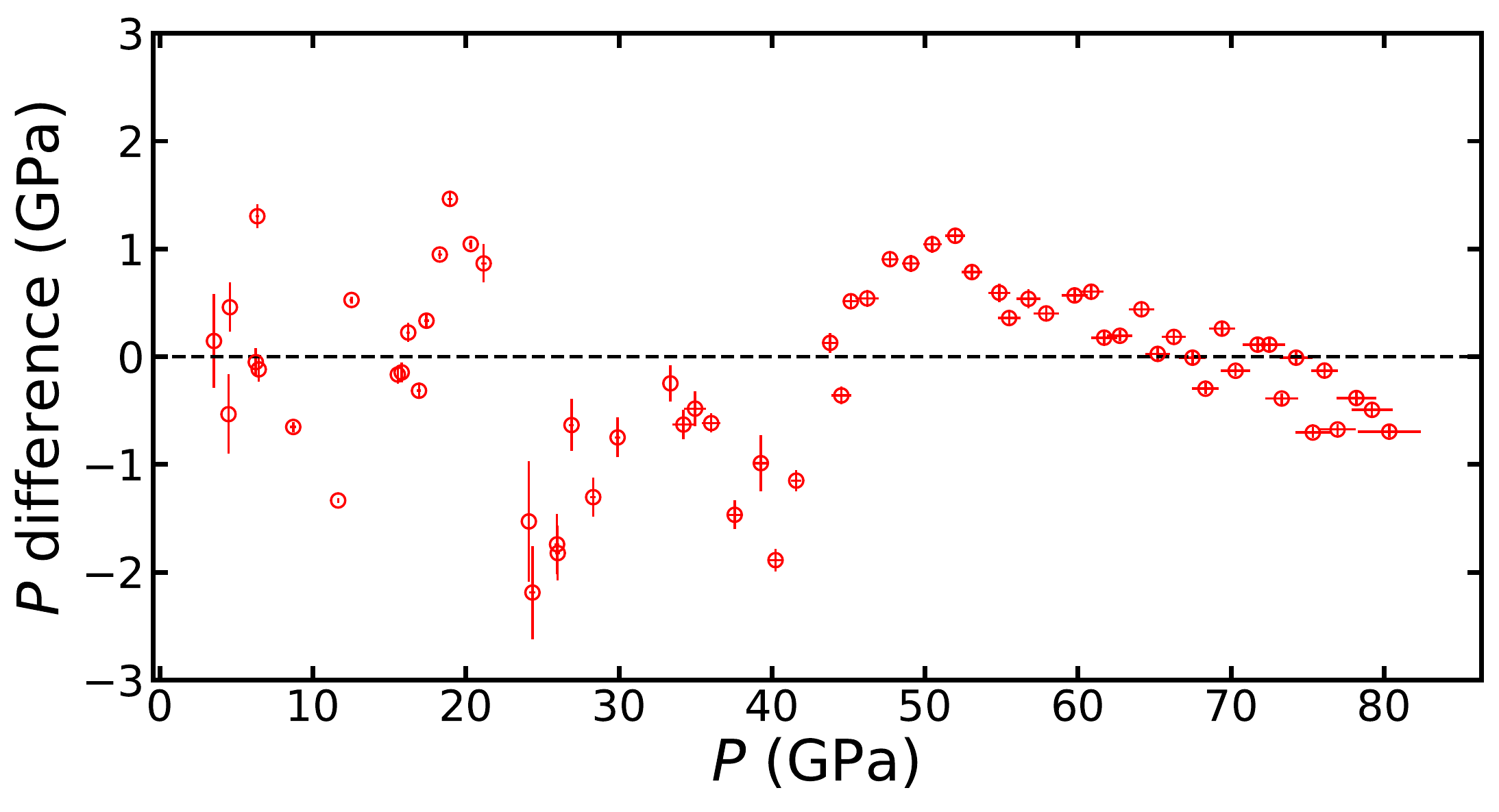}
    \caption{Difference in P between patterns focused on \usix and Cu. For each pair of patterns, the $P$ gradient was estimated based on the shift in the \usix $321$ reflection (the only peak visible in all patterns). The higher uncertainties from 24–34 GPa were due to a decrease in \usix $321$'s relative intensity over that range, which reduced the precision of the peak fitting. The absolute value of each $P$ difference is incorporated into the $P$ error bars shown in Figure 5 in the main text.}
    \label{fig:pgradient}
\end{figure}

Due to its high electron density, \usix is a very strong diffractor. Consequently, \usix peaks overwhelmed those of Cu and Ne in patterns focused on the sample flake (Figure \ref{fig:patternfig}, top). Since several well-defined Cu peaks were desirable for accurate refinement of sample pressures, at each step in membrane pressure we collected an additional diffraction pattern focused on the Cu flake ($\sim$14 \textmu m away from the sample’s focal position). In these patterns, Cu and Ne peaks were more visible (Figure \ref{fig:patternfig}, bottom), allowing us to refine 3 Cu peak locations with Igor Pro. Those peaks’ respective Miller indices ($h$,$k$,$l$) lattice plane spacings ($d_{hkl}$) were fitted with orthogonal distance regression to the line
\begin{equation}
    d_{hkl} = \frac{a}{\sqrt{h^2+k^2+l^2}}
\end{equation}
where $a$ is the unit cell parameter. The unit cell parameter was used to determine $P$ via a Cu EoS \cite{dewaele2004equations}, with $P$ uncertainty deriving from the uncertainty on peak location, fit to Cu unit cell parameter, and EoS parameters. Additional uncertainty is introduced by our assumption that $P$ at the Cu focal point is equal to $P$ at the \usix focal point. It would be desirable to directly compare \usix $V$ between the sample-focused and Cu-focused patterns, but the $321$ peak was the only one observed in all diffraction patterns, and it is impossible to determine $V$ of a tetragonal crystal from a single peak. Therefore, the pressure gradient between the pairs of patterns was estimated by iteratively fitting the EoS parameters assuming no $P$ gradient and comparing the observed displacement of the \usix $321$ peak to the displacement predicted by the EoS. For example, if increasing $P$ by 0.1 GPa was expected to shift peak 321 by 0.1 Å$^{-1}$, an observed shift of -0.3 Å$^{-1}$ was taken to indicate that $P$ was 0.3 GPa lower in the Cu-focused pattern. We then added this $P$ difference to the other sources of $P$ uncertainty and re-fit the EoS. Since estimated $P$ gradients were small and did not systematically vary as $P$ increased (Figure \ref{fig:pgradient}), there was no statistically significant effect on our fit EoS parameters. Additionally, the absence of a systematic trend in $P$ gradient implies that the sample chamber was approximately hydrostatic. 

\begin{figure}[H]
    \centering
    \includegraphics[scale = 0.4]{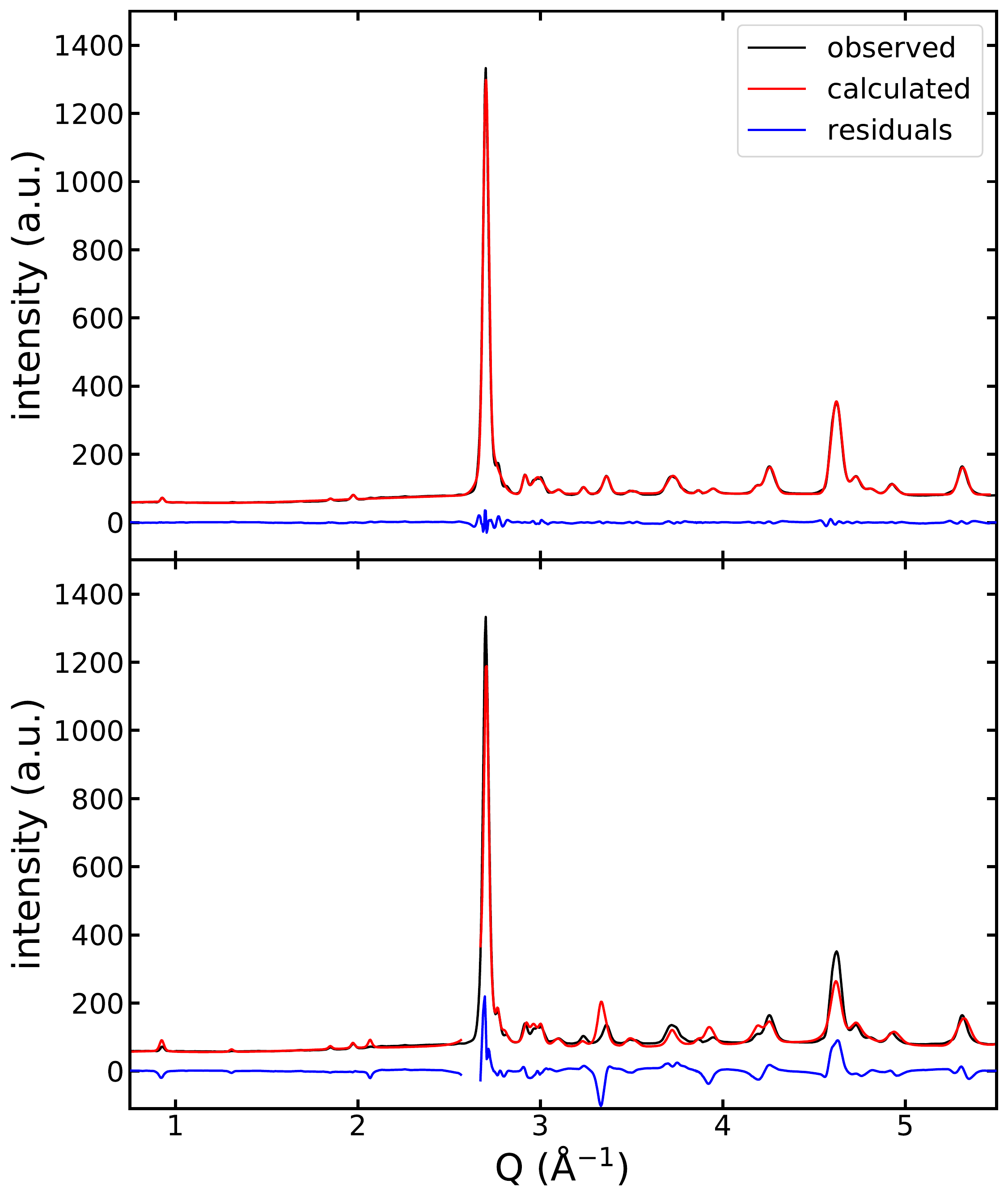}
    \caption{The same integrated diffraction pattern analyzed in Igor Pro by fitting peaks (top) and in GSAS-II by whole-pattern Rietveld refinement (bottom). The region near $Q$ = 2.7 was excluded from the Rietveld refinement to compensate for the absence of \usix $002$.}
    \label{fig:patterncompare}
\end{figure}

\begin{figure}[H]
    \centering
    \includegraphics[scale = 0.5]{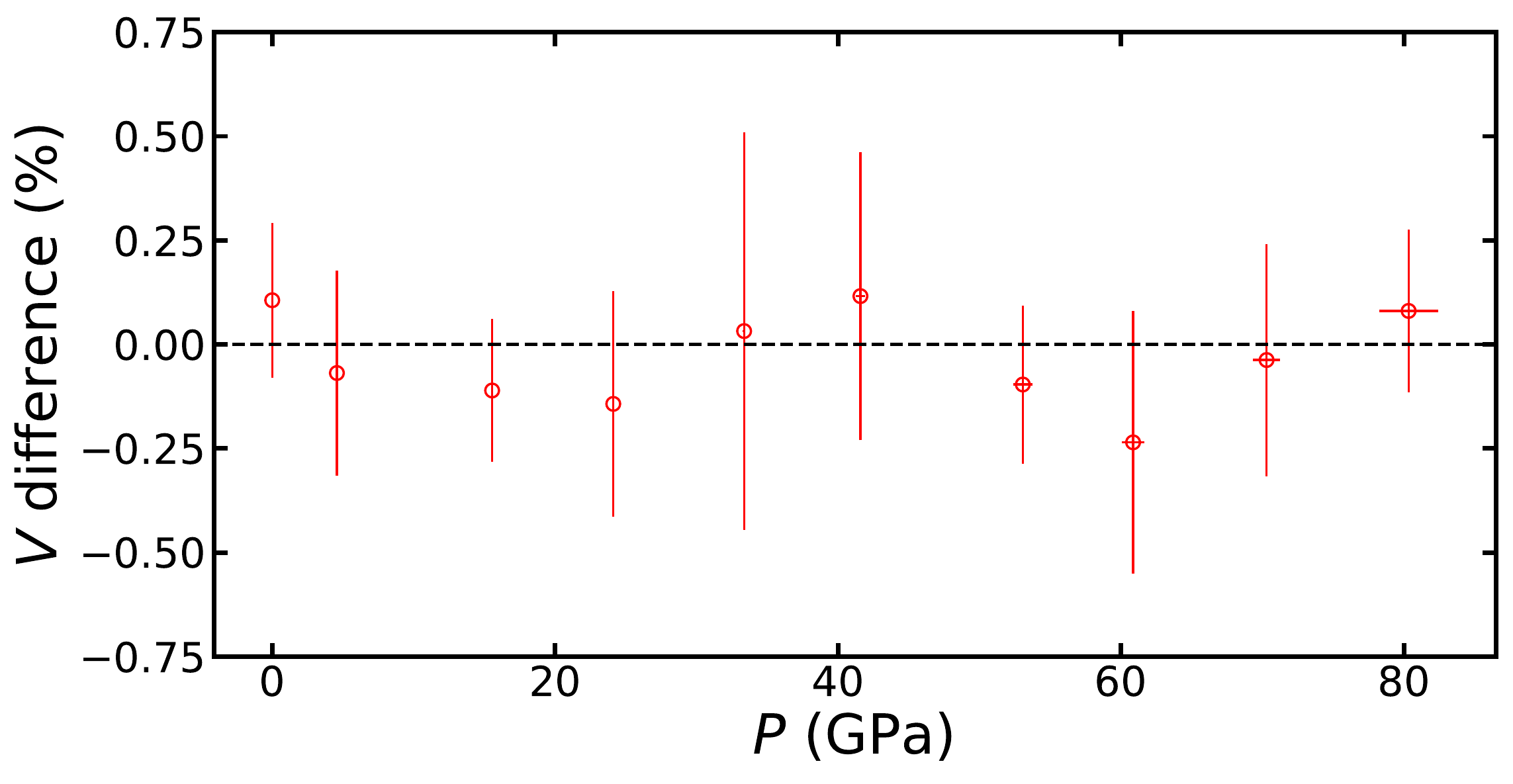}
    \caption{The difference between unit cell $V$ calculated by peak fitting and Rietveld refinement. For all the datapoints tested, calculated unit cell $V$ agreed within uncertainty.  }
    \label{fig:rietveldcompare}
\end{figure}

Sample volumes were determined from the patterns focused on the sample flake by refining an average of 8 \usix peak locations (18 for the ambient-$P$ experiment) with Igor Pro and then using those peaks’ respective $h$,$k$,$l$ and $d_{hkl}$ to perform an orthogonal distance regression fit to the plane
\begin{equation}
    \frac{1}{d_{hkl}^2} = \frac{h^2+k^2}{a^2}+ \frac{l^2}{c^2}
\end{equation}
where $a$ and $c$ are the unit cell parameters. Volume uncertainty derives from uncertainty on the \usix peak locations and the fit to the \usix unit cell. To check the robustness of our unit cell determinations, we performed whole-pattern Rietveld fits with GSAS-II \cite{toby2013gsas} to a subset of the data (Figures \ref{fig:patterncompare},\ref{fig:rietveldcompare}). Calculated unit cells were consistent with the peak-fitting method across our $P$ range (Figure \ref{fig:rietveldcompare}).

\section{\label{sec:level1}Unit cell parameters}

\begin{figure}[h]
    \centering
    \includegraphics[scale = 0.4]{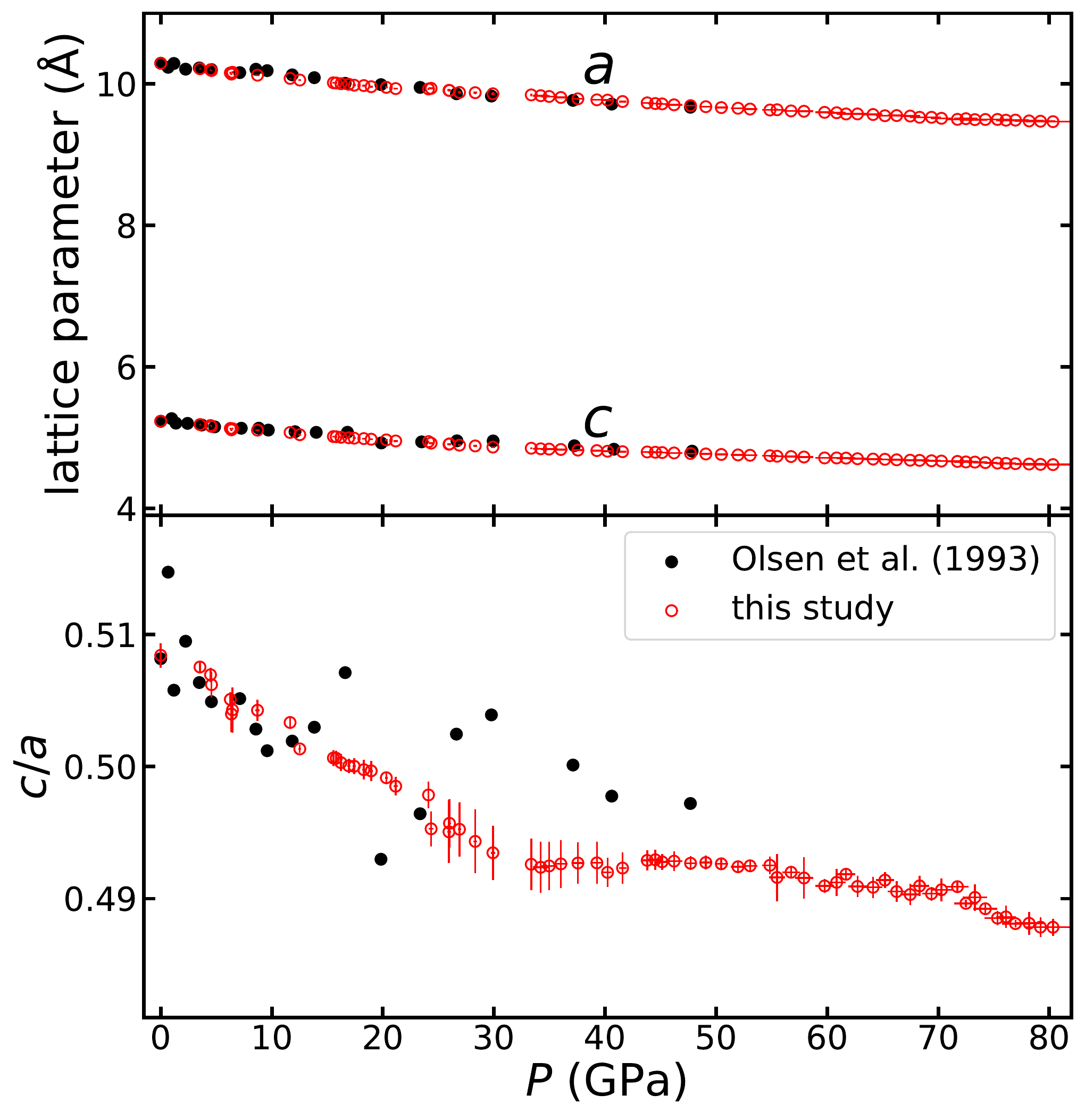}
    \caption{Evolution of unit cell dimensions $a$ and $c$ (top) and the ratio between them (bottom) with increasing $P$.}
    \label{fig:unitcell}
\end{figure}

\section{\label{sec:level1}Olsen et al. (1993)}
\begin{figure}[H]
    \centering
    \includegraphics[scale = 0.5]{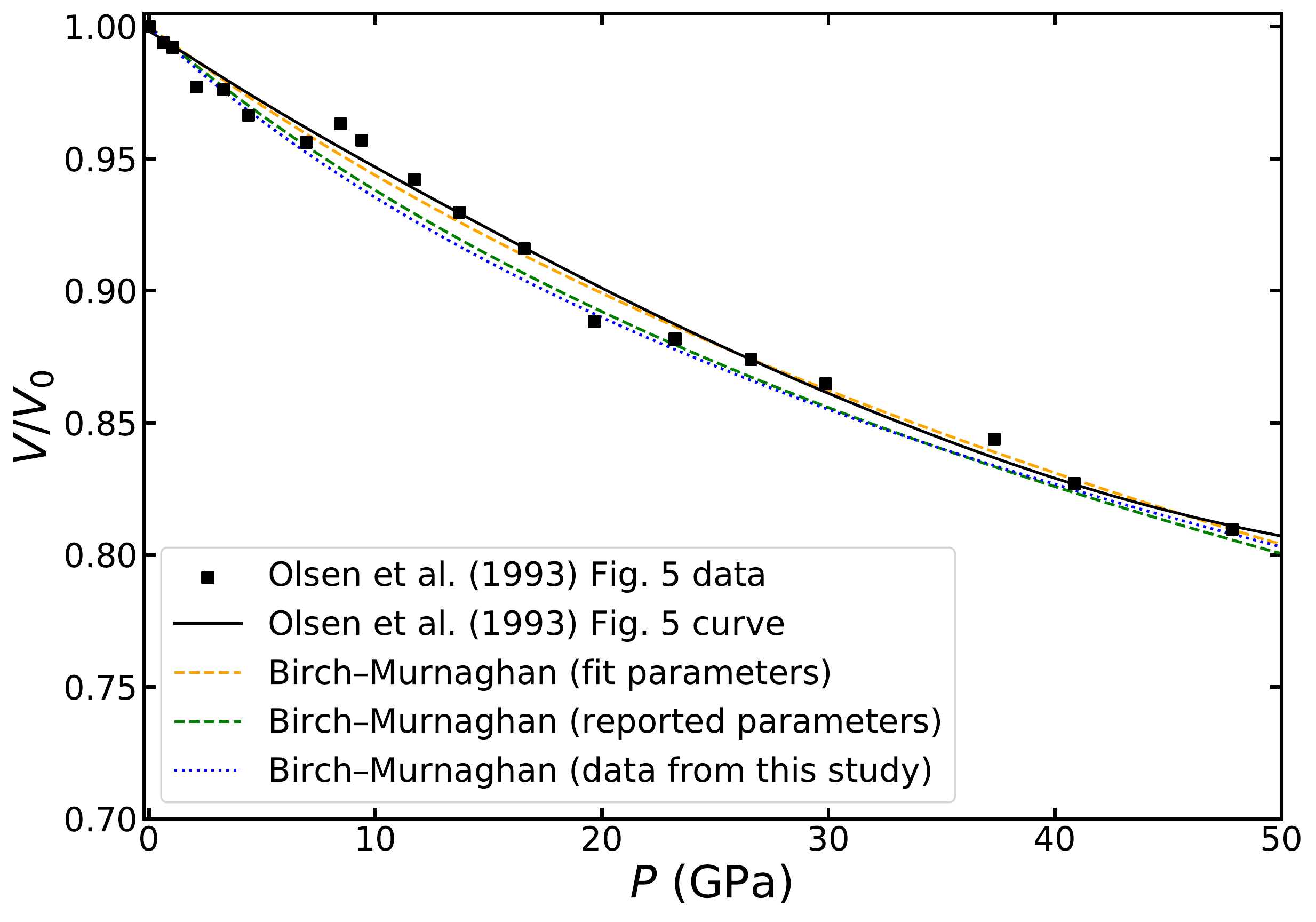}
    \caption{Reproduction of the $V$ versus $P$ figure from Olsen et al. (1993), showing the data and trendline originally displayed, an EoS fit to the data displayed, and a trendline calculated from that study’s reported EoS values. Trendlines using Murnaghan EoS parameters are not shown because they are indistinguishable from their Birch--Murnaghan counterparts. A normalized version of a fit to the data collected in this study (i.e., the ``Birch--Murnaghan fit'' in Figure 5 of the main text) is shown for context. }
    \label{fig:olsenfig}
\end{figure}

Prior to this study, Olsen et al. (1993) \cite{olsen1993developments} was the only publication that reported $P$ versus $V$ data for U\textsubscript{6}Fe. Unfortunately, there is a discrepancy between that study's data and its reported EoS parameters. Fitting the $P$ versus $V$ data shown in Olsen et al. closely reproduces the trendline shown, but calculating an isothermal compression curve from the reported parameters does not (Figure \ref{fig:olsenfig}). Interestingly, the reported parameters are much closer to the results of this study (i.e, Table II in the main text) than the fit parameters. Table \ref{tab:table1} lists Olsen et al.'s reported EoS parameters and the parameters required to fit the data displayed in that study's figure. Comparisons between this study and Olsen et al. in the main text all use the fit, rather than reported, parameters.

\begin{table}
\caption{\label{tab:table1} EoS parameters for the Olsen et al. (1993) fits shown in Figure \ref{fig:olsenfig}, showing the discrepancy between that study’s reported values and the values required to reproduce that study’s figure. All fits use $V_0$ fixed at 555.81 ± 2.89 Å$^3$, as reported. Errors for reported values are as reported, errors for fit values are 1$\sigma$. }
\begin{ruledtabular}
\begin{tabular}{llcc} 
Type of fit & & $B_0$ (GPa) & $B_0'$\\
\colrule
Murnaghan & (reported) & 135.7(50) & 4.25(40)\\
& (fit to data) & 156.9(86) & 3.3(6)\\
\hline
Birch--Murnaghan & (reported) & 134.9(70) & 4.55(60)\\
& (fit to data) & 155.4(85) & 3.6(6)
\end{tabular}
\end{ruledtabular}
\end{table}

\section{\label{sec:level1}SESAME}

\begin{figure}[h]
    \centering
    \includegraphics[scale = 0.4]{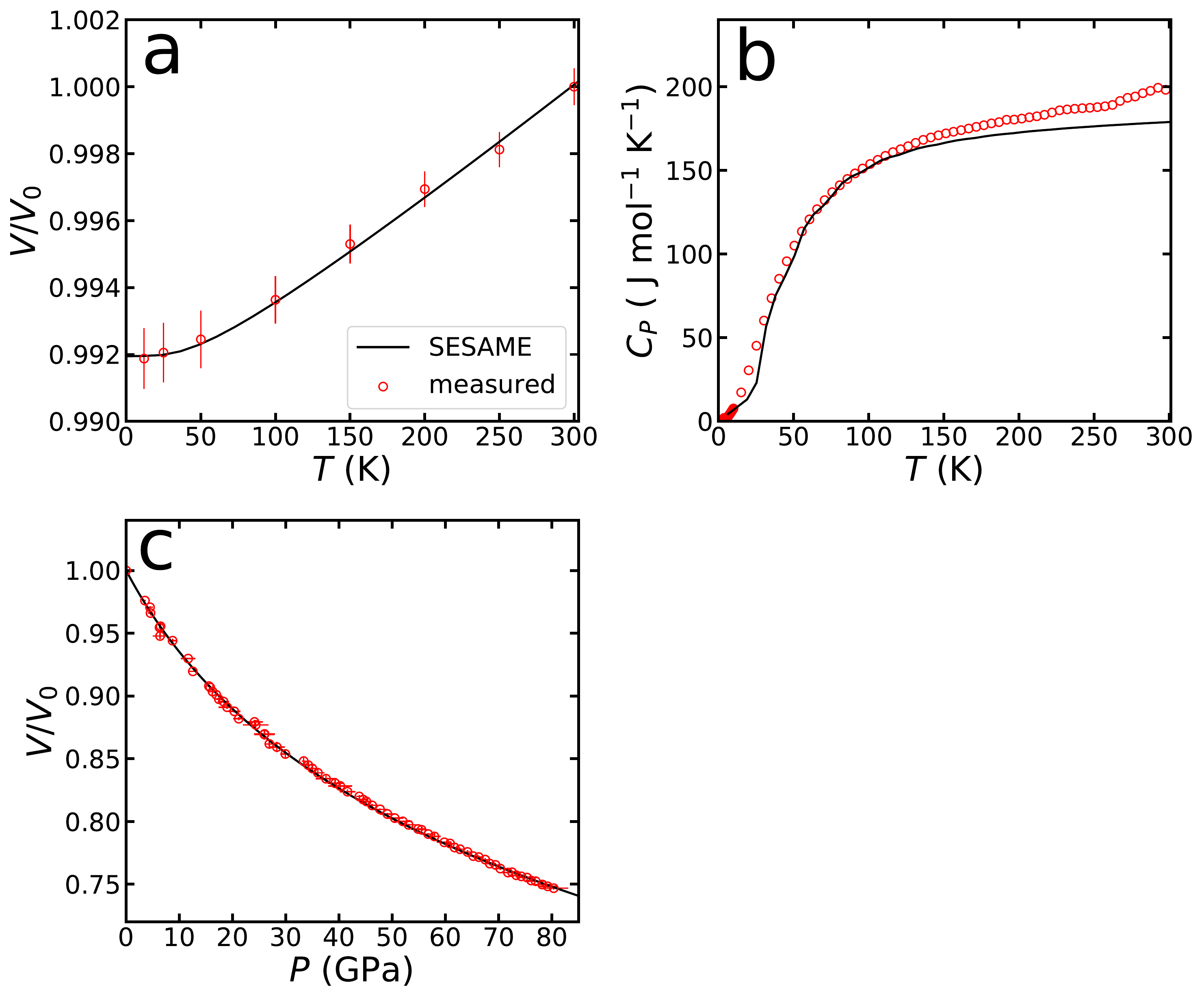}
    \caption{Reproduction of experimental measurements (a: thermal expansivity, b: heat capacity, c: isothermal compression) by SESAME calculations \cite{Lyon1992}.}
    \label{fig:sesame}
\end{figure}

\clearpage

\section{\label{sec:level1}DFT}

\begin{figure}[h]
    \centering
    \includegraphics[scale = 0.5]{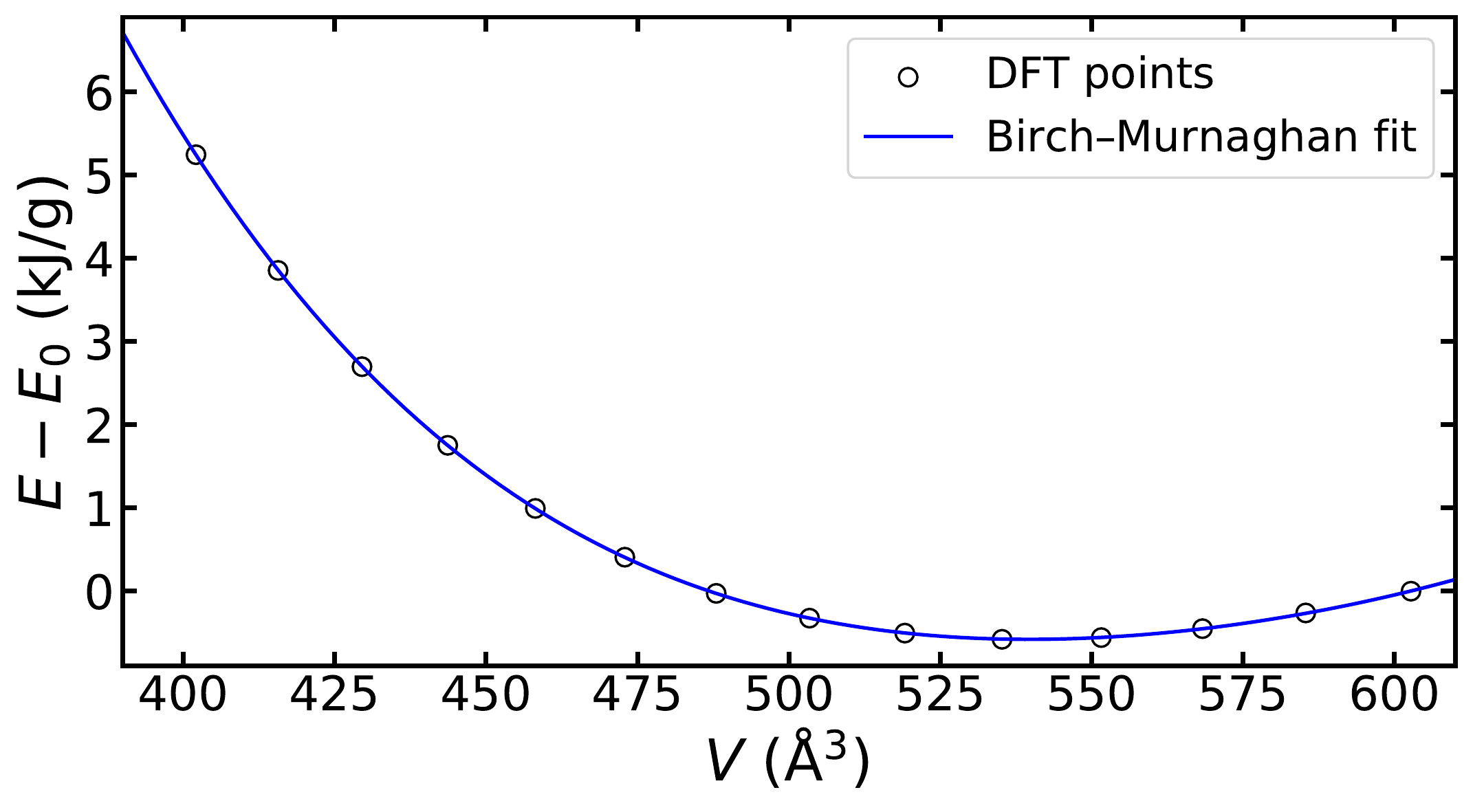}
    \caption{DFT results calculated with VASP \cite{Kresse1993,Kresse1994,Kresse1996a,Kresse1996b} showing the relationship between unit cell volume and internal energy for \usix at 0 K. The fit shown here is described by Eq.~(9) of the main text and was used to determine the DFT EoS parameters shown in Table II of the main text.}
    \label{fig:dft_volume}
\end{figure}

\begin{figure}[h]
    \centering
    \includegraphics[scale = 0.5]{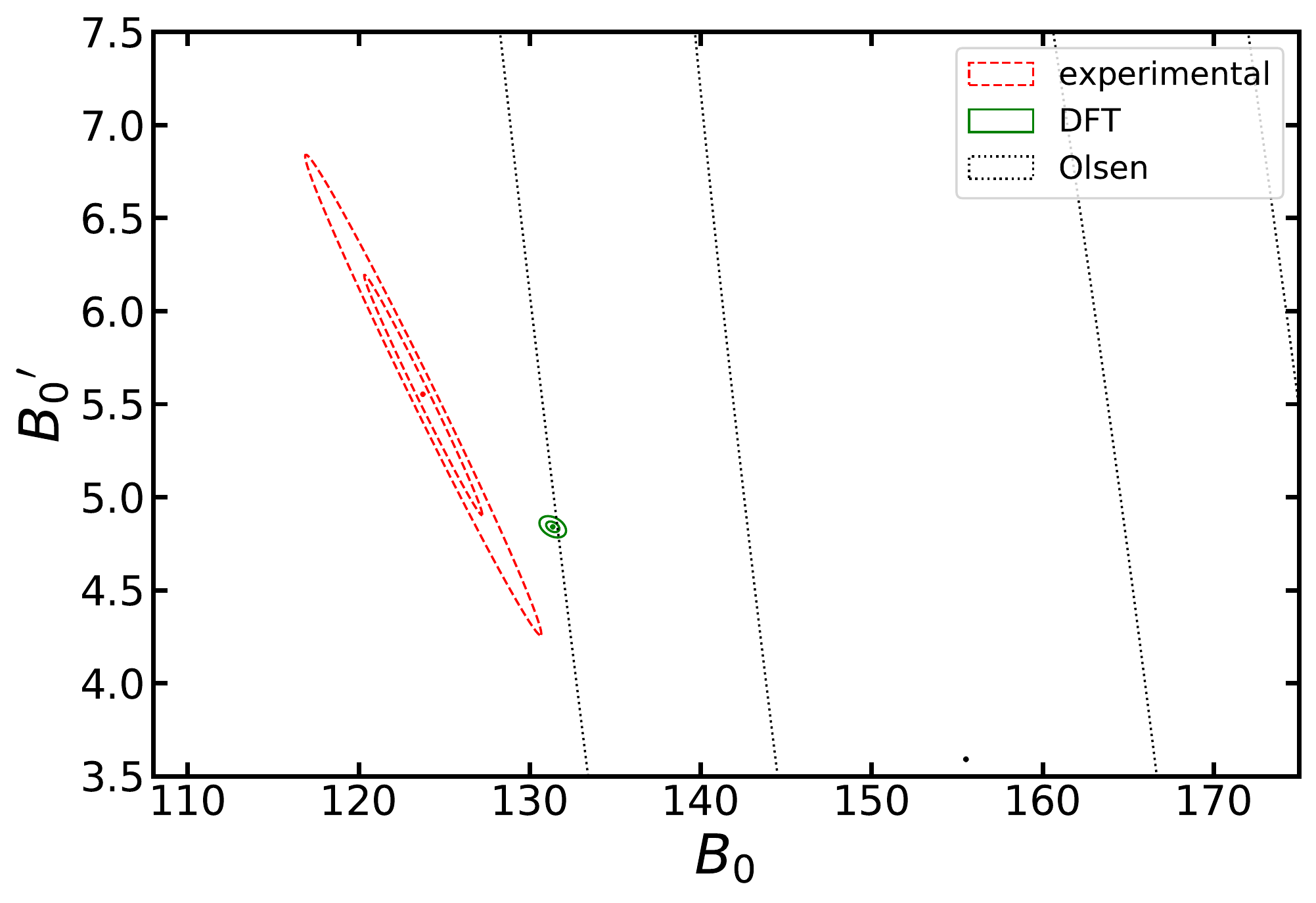}
    \caption{Confidence ellipses showing 1$\sigma$ and 2$\sigma$ covarience between Birch--Murnaghan EoS parameters fit to DFT and experimental data. Covarience values are listed in Table II of the main text. The Olsen et al. (1993) \cite{olsen1993developments} ellipses are not shown in full and extend to unphysical regions (i.e., $B'_0 < 0$) even at the 1$\sigma$ level.}
    \label{fig:covariance}
\end{figure}

\clearpage % force a pagebreak and flush all deferred `table` and `figure` environments
\bibliography{U6Fe_supp_ref.bib}% Produces the bibliography via BibTeX.
%\printbibliography